\documentclass[%
 reprint,
 amsmath,amssymb,
 aps,
]{revtex4-1}

\usepackage{graphicx}
\usepackage{dcolumn}
\usepackage{bm}
\usepackage{upgreek}
\usepackage{color}

\begin{document}

\preprint{APS/123-QED}

\title{Extended coherence length and depth ranging using a Fourier domain mode-locked frequency comb and circular interferometric ranging}

\author{Norman Lippok$^{1,2}$} \email{nlippok@mgh.harvard.edu}
\author{Meena Siddiqui$^{1,2}$}
\author{Benjamin J. Vakoc$^{1,2,3}$}
\author{Brett E. Bouma$^{1,2,3}$} 

\affiliation{
 $^1$Harvard Medical School, Boston, MA 02115, USA\\
 $^2$Wellman Center for Photomedicine, Massachusetts General Hospital, Boston, MA 02114, USA\\
 $^3$Institute for Medical Engineering and Science, Massachusetts Institute of Technology, Cambridge, MA 02139, USA
}




\date{\today}

\begin{abstract}
Fourier domain mode-locking (FDML) has been a popular laser design for high speed optical frequency domain imaging (OFDI) but achieving long coherence lengths, and therefore imaging range, has been challenging. The narrow instantaneous linewidth of a frequency comb (FC) FDML laser could provide an attractive platform for high speed as well as long range OFDI. Unfortunately, aliasing artifacts arising from signals beyond the principle measurement depth of the free spectral range have prohibited the use of a FC FDML for imaging so far. To make the enhanced coherence length of FC FDML laser available, methods to manage such artifacts are required. Recently, coherent circular ranging has been demonstrated that uses frequency combs for imaging in much reduced RF bandwidths. Here, we revisit circular ranging as a tool of making the long coherence length of an FDML frequency comb laser as well as its use for tissue imaging accessible. Using an acousto-optic frequency shifter (AOFS), we describe an active method to mitigate signal aliasing that is both stable and wavelength independent. We show that an FC FDML laser offers an order of magnitude improved coherence length compared to traditional FDML laser designs without requiring precise dispersion engineering. We discuss design parameters of a frequency stepping laser resonator as well as aliasing from a frequency comb and AOFS in OFDI with numerical simulations. The use of circular ranging additionally reduced acquisition bandwidths 15-fold compared with traditional OFDI methods. The FC FDML/AOFS design offers a convenient platform for long range and high speed imaging as well as exploring signal and image processing methods in circular ranging.
\end{abstract}

\maketitle


\begin{figure*}[t]
\centerline{\includegraphics[width=\linewidth]{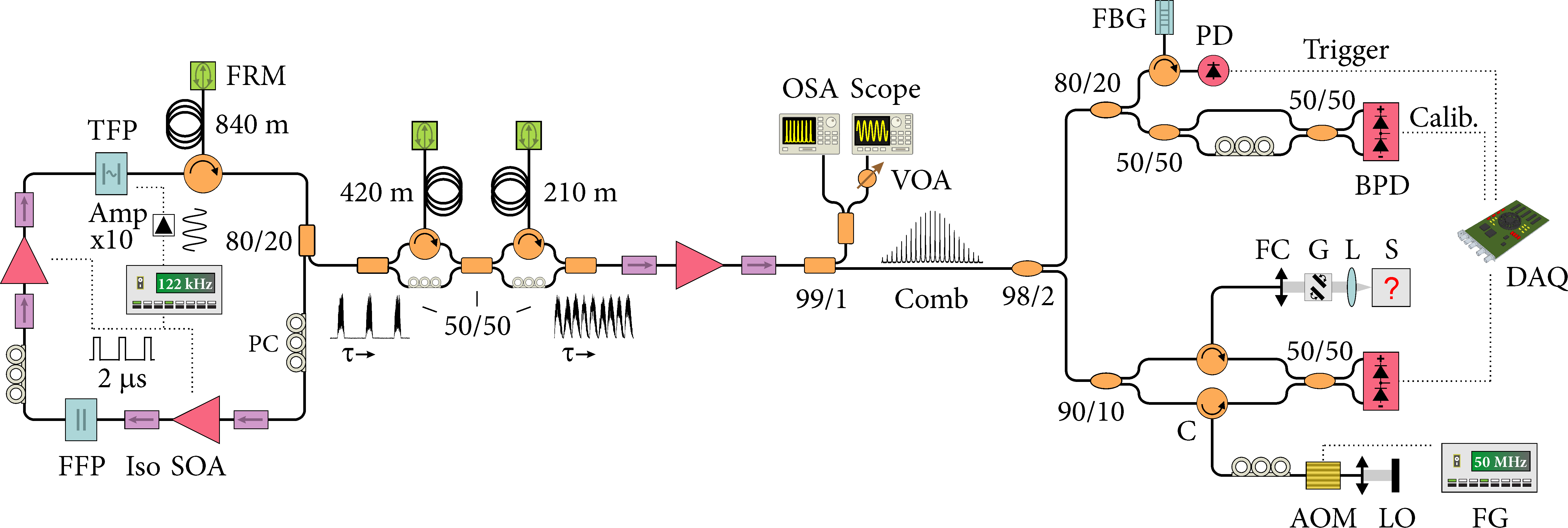}}
\caption{\label{Fig1} Experimental setup. FFP, fixed Fabry-P\'erot etalon; TFP, tunable Fabry-P\'erot etalon; Iso, isolator; SOA, semiconductor optical amplifier; PC, polarization controller; FRM, Faraday rotating mirror; VOA, variable optical attenuator; OSA, optical spectrum analyser; FBG, fiber Bragg grating; PD, photo diode; BPD, balanced photo diode; C, circulator; FC, fiber collimator; G, 2D Galvanometer mirror set; L, objective lens; S, sample; AOM, acousto-optic modulator; LO, local oscillator; FG, frequency generator.}
\end{figure*}

\section{Introduction}

Wavelength swept laser sources used for optical frequency domain imaging (OFDI) have been the focus of intense research to improve their repetition rate and coherence length over the past decade~\cite{Yun,Huber,Fuji:1}. The Fourier domain mode-locked (FDML) laser has become increasingly attractive for high speed imaging~\cite{Huber,Adler,Wieser:3,Wieser:1,Klein,Biedermann,Pfeiffer}. It uses a long optical fiber delay line and a fiber Fabry-P\'erot tunable filter (TFP) with a sweep rate that is synchronized with the roundtrip time of the cavity, thereby improving output power and tuning rate. However, its depth range has been limited to only a few millimeters. Current strategies to extend depth range in FDML rely on precise dispersion engineering of the cavity using a set of chirped, broadband optical fiber Bragg gratings that have provided marginal extensions of imaging range to 5~mm and 1~cm at 1.6 MHz (30 THz/$\upmu$s) and 200~kHz (3.6 THz/$\upmu$s) repetition rate (sweep speed), respectively~\cite{Adler,Wieser:3}. Recently, extremely precise (very high order) dispersion control was implemented by applying a temperature gradient (with $\sim$ 0.1$^o$C resolution) along a fiber Bragg grating, while simultaneously managing temperature of the gain medium, tunable spectral filter and housing, achieving imaging ranges of 10 cm at 3 MHz repetition rate and 77 nm sweep bandwidth (40 THz/$\upmu$s) for the first time~\cite{Pfeiffer}. While impressive in performance, this yields environmentally sensitive laser cavities and asks for more than 100 GS/s acquisition bandwidths, which is not practical. Frequency combs (FC) can provide an alternative strategy to increase coherence length. FC FDML has been proposed in the past but the FC induced a degeneracy in the ranging measurements that resulted in imaging artifacts. These artifacts effectively limited the imaging range to a few millimeters, and thus no significant improvement was achieved~\cite{Tsai}.\\
\indent Recently, circular ranging optical coherence tomography has been introduced to reduce bandwidth requirements without compromising range, speed, or axial resolution~\cite{Siddiqui:0,Siddiqui}. In this approach, the depth space of the interferometric signals is folded, leveraging sparsity within the imaged field to reduce the number of measurements needed to capture the sample properties. Circular ranging is implemented in the optical domain by combining a frequency comb source with an interferometric system capable of discriminating between positive and negative differential delays. While this technique has been proposed for the primary purpose of bandwidth reduction, it also provides a means for overcoming degeneracy artifacts induced by using a FC source. In this work, we demonstrate that these methods from circular ranging can be adopted to avoid artifacts that have limited FC FDML lasers. Using an acousto-optic frequency shifter (AOFS), we discriminate positive from negative delay and mitigate aliasing signals outside the principle measurement range, making the long coherence length of FC FDML accessible and its application for tissue imaging possible.\\
\indent The combination of FC FDML and circular ranging offers one order of magnitude improved coherence length compared to standard FDML sources~\cite{Huber,Wieser:1}, without the temperature control and accurate high order dispersion compensation that has previously been required. We demonstrate imaging results obtained over centimetre-ranges at 488~kHz repetition rate and 105 nm (18 THz) wavelength sweep range.  In addition, the use of circular ranging reduced acquisition bandwidths 15-fold compared with traditional OFDI methods. Moreover, the active AOFS approach is stable and wavelength independent, yielding an advantage over the previously proposed passive polarization demodulation method used for circular ranging~\cite{Siddiqui:0,Siddiqui:2}. We provide new insight for the design parameters of a frequency stepped laser and discuss aliasing signals arising from a frequency comb source in OFDI using numerical simulations.\\
\indent This FDML/AOFS circular ranging architecture offers high speed, long range imaging and provides a convenient and accessible platform for developing and optimizing the novel and unique aspects of circular ranging. In addition, it demonstrates a broader strategy for converting existing swept-wavelength laser designs to operate as stepped frequency comb sources for circular ranging to overcome signal acquisition limitations. 

\section{FDML frequency comb laser}\label{Sec2}

\begin{figure}[!t]
\centerline{\includegraphics[width=6.8cm]{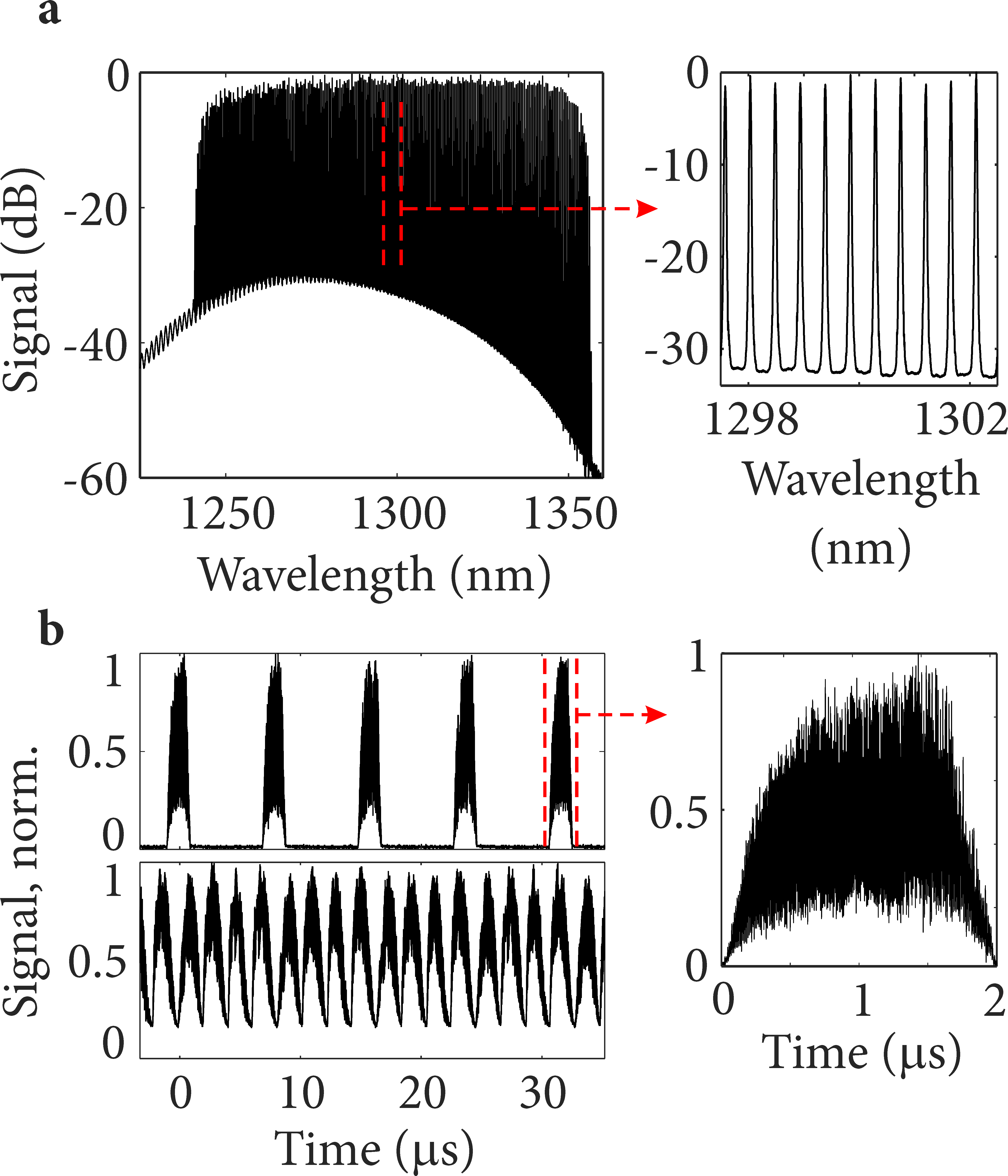}}
\caption{(a) Experimental spectrum showing a frequency comb spanning 18.5~THz with a line spacing of 80~GHz. (b) Time traces showing the laser output at 25~\% duty-cycle (top) and 100~\% duty-cycle (bottom). A magnified view illustrates strong pulsation within the slowly varying envelop of a single TFP sweep.}
\label{Fig2}
\end{figure}

\begin{figure}[!t]
\centerline{\includegraphics[width=7.7cm]{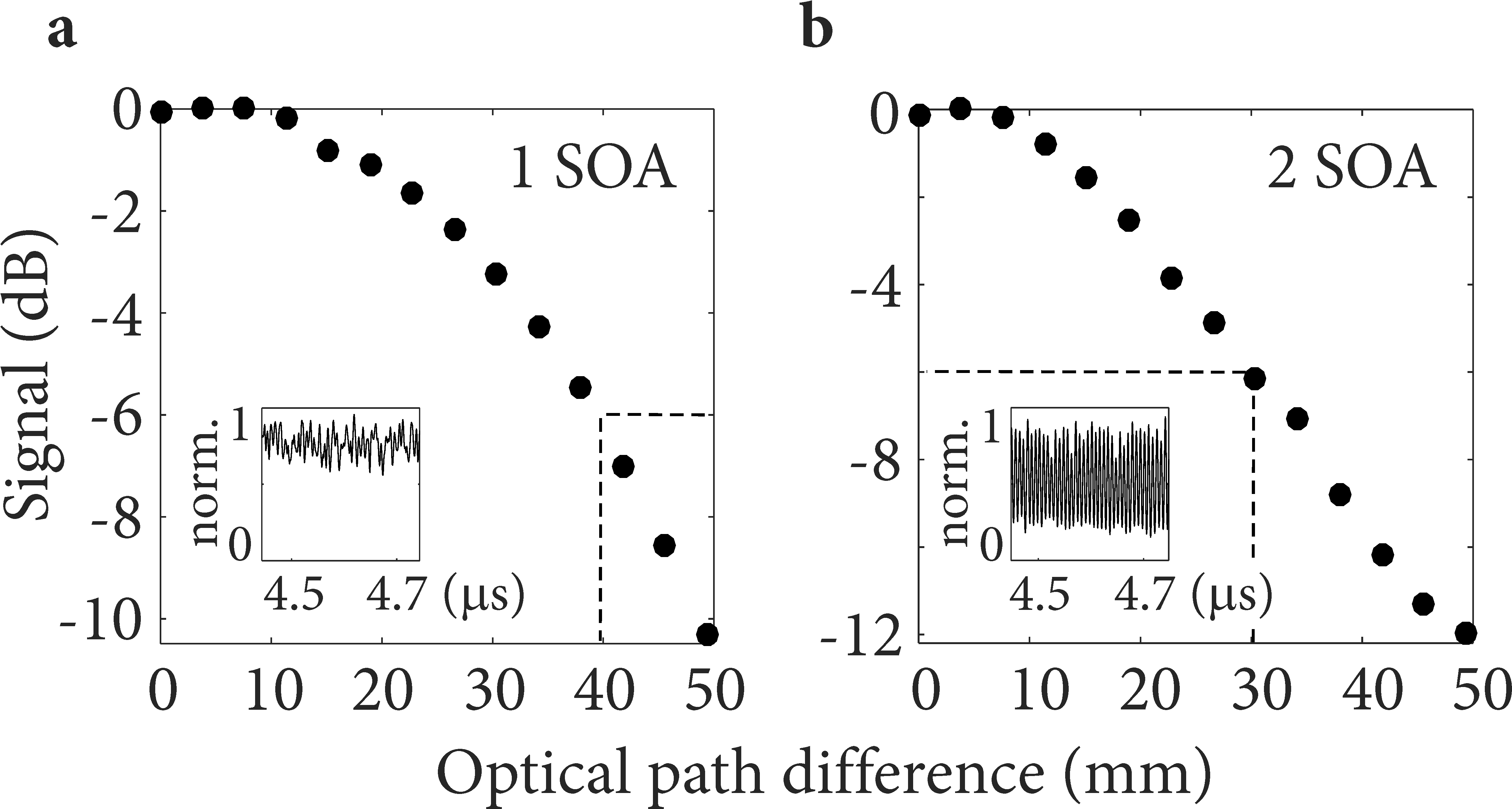}}
\caption{Measured coherence length for a single SOA (a) and two SOAs (b) inside the cavity. The insets in (a) and (b) show magnified time traces.}
\label{Fig3}
\end{figure}

By incorporating an intracavity Fabry-P\'erot etalon into the extended cavity of an FDML laser, the nested cavity produced a frequency comb that was superimposed onto the FDML longitudinal modes (frequency comb is independent from the FDML cavity) (Fig.~\ref{Fig1}). The 1.68~km fiber (Corning SMF28) resonator included a 1310~$\pm$~40~nm semiconductor optical amplifier (SOA) (Covega, BOA-7875) as gain medium, a fixed Fabry-P\'erot etalon (FFP, LightMachinery, Inc.) with a free spectral range (FSR) of $\Delta\nu_{\mathrm{FFP}}$ = 80~GHz and a finesse of 80 and a tunable Fabry-P\'erot filter (TFP, Micron Optics, Inc.) with an FSR of $\Delta\nu_{\mathrm{TFP}}$ = 36~THz and a finesse of $\sim$900. The TFP was driven at the resonator's fundamental resonance of 122~kHz. This frequency was closely matched to the second harmonic of the TFP mechanical resonance frequency. The SOA was current modulated, using a 2~$\upmu$s electrical pulse, to suppress lasing during the long to short wavelength (backward) sweep, and to enable lasing for 50\% of each forward sweep. This resulted in a lasing bandwidth of 105 nm (18~THz) and an average sweep speed of 52~nm/$\upmu$s (9~THz/$\upmu$s). In some configurations, a second SOA was included to increase output power and reduce laser noise. The laser output was directed through delay lines to generate four time-shifted copies prior to amplification. This increased the duty cycle from ~25\% to nearly 100\% and the repetition rate from 122 kHz to 488 kHz.\\
\indent Figure~\ref{Fig2} shows the laser output in the optical frequency and time domains. The frequency comb source extended across a 105~nm (18.5~THz) bandwidth. Individual comb lines are visible in a magnified view showing a line spacing of 80~GHz ($\sim$0.45~nm). The time-trace shows five consecutive sweeps at 25\% duty cycle obtained directly at the laser output [Fig.~\ref{Fig2}(b), top] and a full duty cycle with 488~kHz repetition rate after the delay line and amplification [Fig.~\ref{Fig2}(b), bottom]. A magnified view depicts a single sweep. Pulsation is clearly visible, and each pulse corresponds to a single frequency comb line. Near the center wavelength of the sweep, the time between two consecutive pulses was 7.2~ns, reflecting the wavelength sweeping speed and fixed etalon FSR as $\Delta t=\Delta\nu_{\mathrm{FFP}}/$\emph{v}, where \emph{v} is sweeping speed of the light source (in Hz/s). The measured pulse width was 4.5~ns with a modulation depth of approximately 0.6. The pulsation induced by the FFP can be used to self-clock acquisition at the optical Nyquist rate (i.e., one digitization per frequency comb line).\\
\indent The measured coherence length was 4~cm as shown in Fig.~\ref{Fig3}(a). Each data point was recorded at integer multiples of the baseband length, $L_B = c/(2\Delta\nu_{\mathrm{FFP}}$), to avoid confounding coherence length with cyclic variations in signal strength across the baseband. The coherence length was smaller than that predicted by the fixed FFP transmission peak linewidth ($\sim$10~cm) due to nonlinear linewidth broadening in the SOA. Figure~\ref{Fig3}(b) shows the measured coherence length when employing two SOAs in the cavity. As expected, the additional nonlinear interactions in the second SOA reduces the coherence length. However, we observed that two SOAs reduced laser noise. This can also be appreciated from the two insets in Fig.~\ref{Fig3}(a),(b) where the pulsation is clearly visible in the case when two SOAs were used. 

\section{Simulations}\label{Sec3}

\begin{figure}[!t]
\centerline{\includegraphics[width=\linewidth]{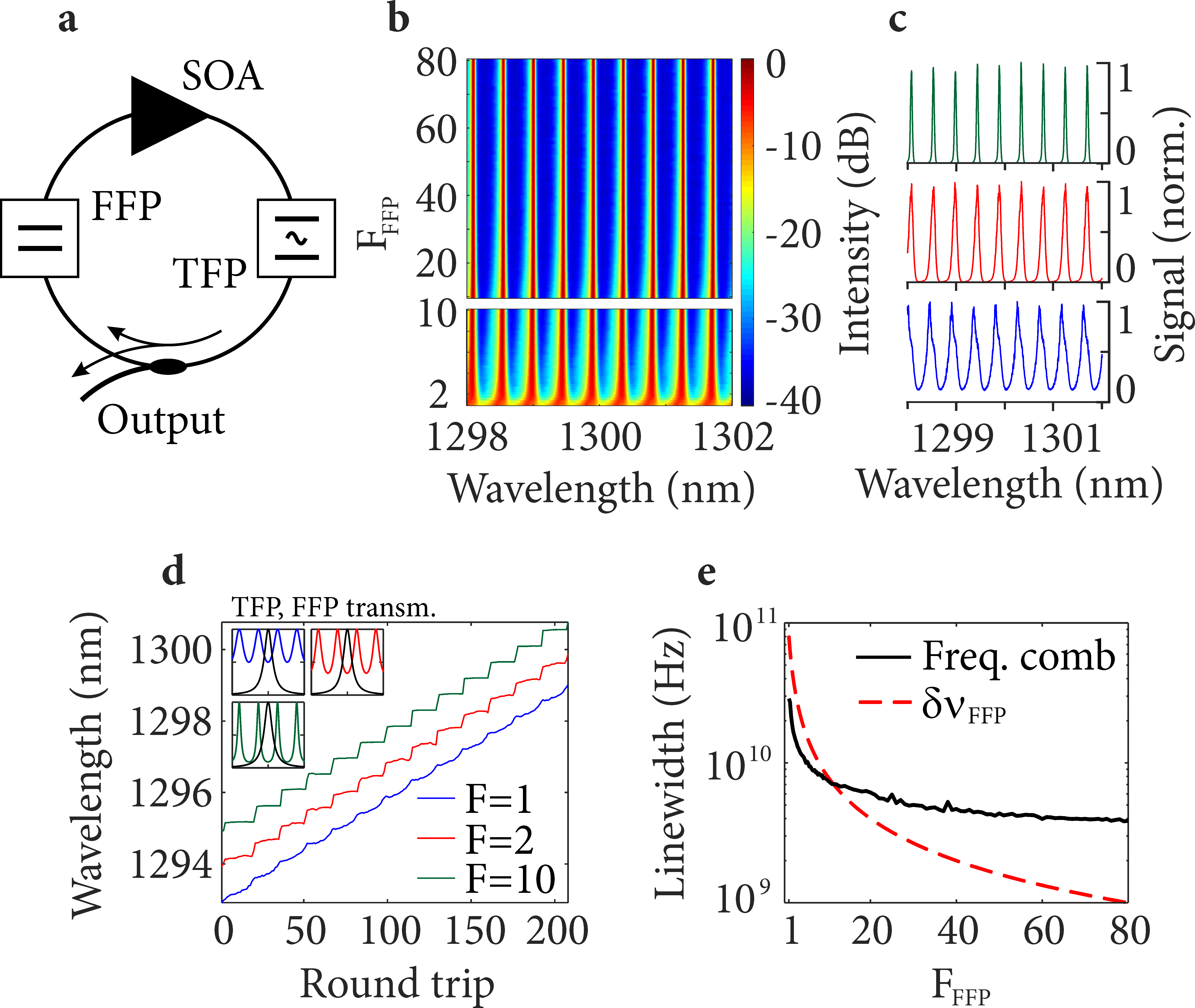}}
\caption{(a) Schematic of a frequency stepped, external cavity, frequency comb laser used for simulations . (b) Simulated optical spectra of the stepped frequency comb laser output with different values of the FFP finesse. (c) Three selected optical spectra for a FFP finesse of 1 (blue), 2 (red) and 10 (green). (d) Simulated instantaneous wavelength per roundtrip (1 round trip = 15 ns) at the laser output for three values of the FFP finesse. Insets in (d) show linear plots of the FFP and TFP (FWHM = 40 GHz ) transmission profiles (spanning $\pm 150$ GHz ) for $F_{\rm{FFP}}=$1, 2, 10. (e) Simulated frequency comb instantaneous linewidth at the laser output for increasing FFP finesse (solid black line). The red, dashed line depicts the FFP bandwidth. FFP FSR was 80 GHz, TFP linewidth was 40 GHz, sweep speed was 2 nm/$\upmu$s.}
\label{Fig4}
\end{figure}

In this section, we discuss design parameters of a stepped frequency comb by considering an extended-cavity semiconductor laser with numerical simulations. We consider a fiber ring cavity that includes an SOA and intracavity TFP and FFP filters [Fig.~\ref{Fig4}(a)]. For simplicity, the numerical model simulates laser performance with short cavity round trip times. A detailed description can be found in Appendix A. Laser parameters are selected partly based on imaging requirements and include choosing a FFP FSR ($\Delta\nu_{\rm{FFP}}$ = 80~GHz) to define a circular depth range ($L_B$ = 1.88 mm). We begin by choosing a TFP bandwidth of $\delta\nu_{\rm{TFP}}$ = 40 GHz, yielding a filter ratio of $\delta\nu_{\rm{TFP}}/\Delta\nu_{\rm{FFP}}$ = 0.5. For circular ranging, a frequency comb with narrow linewidths provides long coherence lengths and correspondingly long depth ranges. The optical spectrum of a frequency comb at the laser output is simulated in Figure~\ref{Fig4}(b) for increasing FFP finesse. Figure~\ref{Fig4}(c) shows three frequency combs in linear scale in more detail for a finesse of 1 (blue), 2 (red) and 10 (green). While only a qualitative assessment, a finesse above 10 yields frequency combs with linewidths that are substantially narrower than the FSR and with high extinction between comb lines. For circular ranging, it is important that the laser wavelength steps discretely between comb lines, and that there is minimal time (if any) in which the laser wavelength oscillates back and forth between two lines. In Fig.~\ref{Fig4}(d) we show the simulated instantaneous wavelength during a sweep as a function of cavity round trips. A finesse of 1 leads to a semi-continuous sweep similar to standard wavelength swept sources, despite the presence of a periodic frequency comb in the spectral domain [Fig.~\ref{Fig4}(d), blue]. A finesse of 2 yields quasi-frequency stepping, where neighboring wavelengths appear to compete for gain, which could lead to artifacts and reduced coherence length [Fig.~\ref{Fig4}(d), red]. A sufficiently high finesse yields frequency stepping with a nearly constant wavelength until discretely switching to the next comb line [Fig.~\ref{Fig4}(d), green]. Figure~\ref{Fig4}(e) shows the frequency comb linewidth at the laser output as a function of FFT finesse. Note that the output linewidth is a function of, but does not directly match, the FFP linewidth. Specifically, for low finesse, linewidths narrower than the FFP bandwidth ($\delta\nu_{\rm{FFP}}$) are observed, while for high finesse, linewidths broader than $\delta\nu_{\rm{FFP}}$ (red, dashed line) occur. Interestingly, while a high finesse is important for frequency stepping, we observed increased laser noise when $\delta\nu_{\rm{FFP}}$ becomes too narrow. The region where linewidths are larger than $\delta\nu_{\rm{FFP}}$ seen in Fig.~\ref{Fig4}(e) could indicate a transition into this noisy regime. This suggests that optimum performance may be achieved at more moderate finesse values, where the narrow FFP bandwidth does not fight with nonlinear spectral broadening in the SOA.

\begin{figure}[!t]
\centerline{\includegraphics[width=8cm]{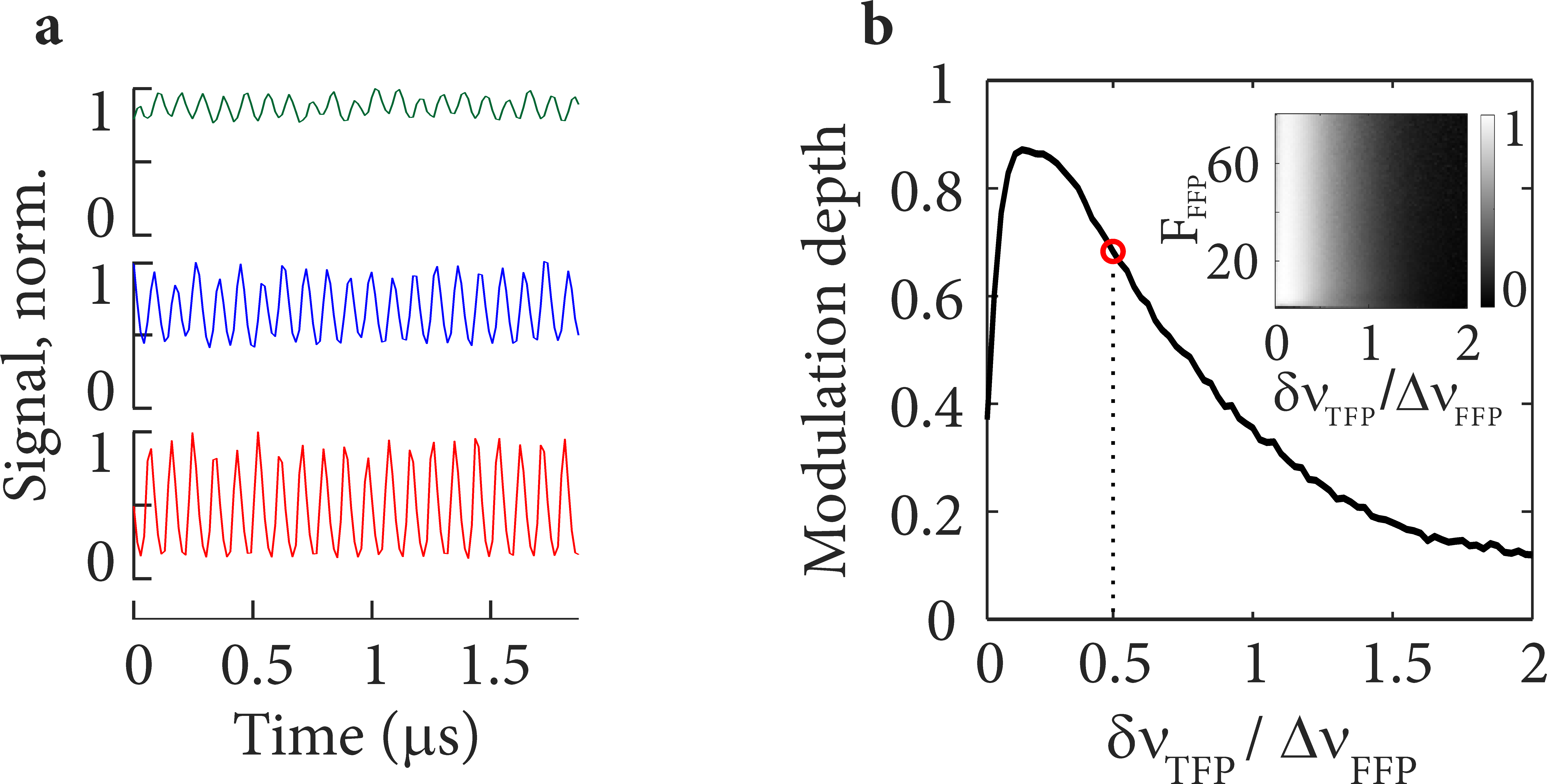}}
\caption{(a) Simulated time traces for selected ratios, $\delta\nu_{\rm{TFP}}/\Delta\nu_{\rm{FFP}}$, of 0.5 (red), 1 (blue) and 2 (green) between the bandwidth of the TFP ($\delta\nu_{\mathrm{TFP}}$) and the FSR of the FFP ($\Delta\nu_{\mathrm{FFP}}$). Pulsation is clearly visible. (b) Simulated pulse modulation depth as a function of filter ratio. The inset shows pulse modulation depth for the same filter ratios versus finesse of the fixed Fabry-P\'erot (FFP). The red circle shows the filter ratio used in our experiments.}
\label{Fig5}
\end{figure}

The TFP bandwidth ($\delta\nu_{\rm{TFP}}$) should be selected to avoid simultaneous lasing of multiple comb lines. Note that as the TFP bandwidth narrows, amplitude pulsation at the output of a stepped frequency comb laser source becomes more pronounced. Figure~\ref{Fig5}(a) shows time traces of the laser output for different ratios between TFP bandwidth, $\delta\nu_{\mathrm{TFP}}$, and FFP FSR, $\Delta\nu_{\mathrm{FFP}}$. The FSR (and finesse) of the FFP was kept constant, while the bandwidth of the TFP was increased. The laser output shows strong pulsation for low filter ratios (small $\delta\nu_{\mathrm{TFP}}$). Modulation depth visibly reduces and transitions into a cw regime for an increasing filter ratio (increasing $\delta\nu_{\mathrm{TFP}}$), which is clearly illustrated in Fig.~\ref{Fig5}(b). Of note, the modulation depth is not influenced by the FFP finesse for $F_{\rm{FFP}}>2$ [Fig.~\ref{Fig5}(b), inset]. For large TFP bandwidths ($\delta\nu_{\rm{TFP}}/\Delta\nu_{\rm{FFP}}>1$), a lasing comb line competes with neighboring comb lines and again challenges the required wavelength stepping shown in Fig.~\ref{Fig4}(d). By adjusting the TFP bandwidth with respect to FFP FSR, the pulsation amplitude can be controlled, which is relevant for self-clocking and determines the RF frequency content of signal harmonics described in section~\ref{Sec4}. 

\section{Artifact suppression in frequency shifted circular ranging}\label{Sec4}

\begin{figure}[!t]
\centerline{\includegraphics[width=7.9cm]{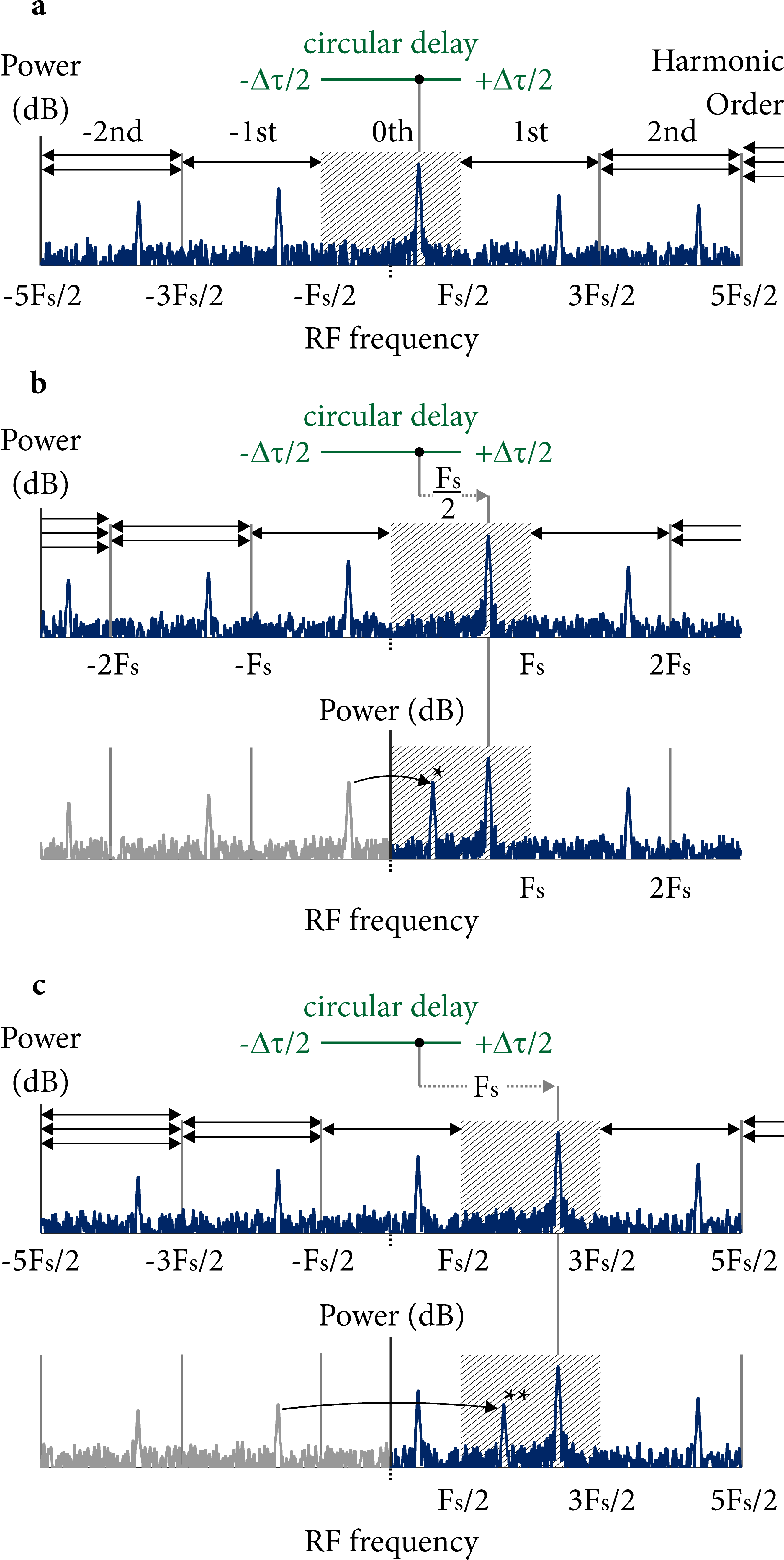}}
\caption{RF harmonic generation and resulting artifacts in time-stepped frequency comb circular ranging. (a) The RF content generated by a single reflection with denoted circular delay is shown for I/Q demodulation and without an AOFS. The signals are located in the baseband centered at DC, and RF harmonics of varying orders occur at higher positive and negative RF frequencies. The amplitude of the RF harmonics drops with order and is a function of the pulse shape. (b) Including an AOFS at Fs/2, while retaining I/Q outputs, shifts the baseband signals to the positive RF frequency space (upper panel). However, using a single channel (I only) causes the negative first order harmonic to appear within the baseband (lower panel denoted by *). (c) Increasing the AOFS frequency shift to Fs places the baseband further from DC, and shifts artifacts to the second negative order which have lower amplitudes.}
\label{Fig6}
\end{figure}

\begin{figure}[!t]
\centerline{\includegraphics[width=\linewidth]{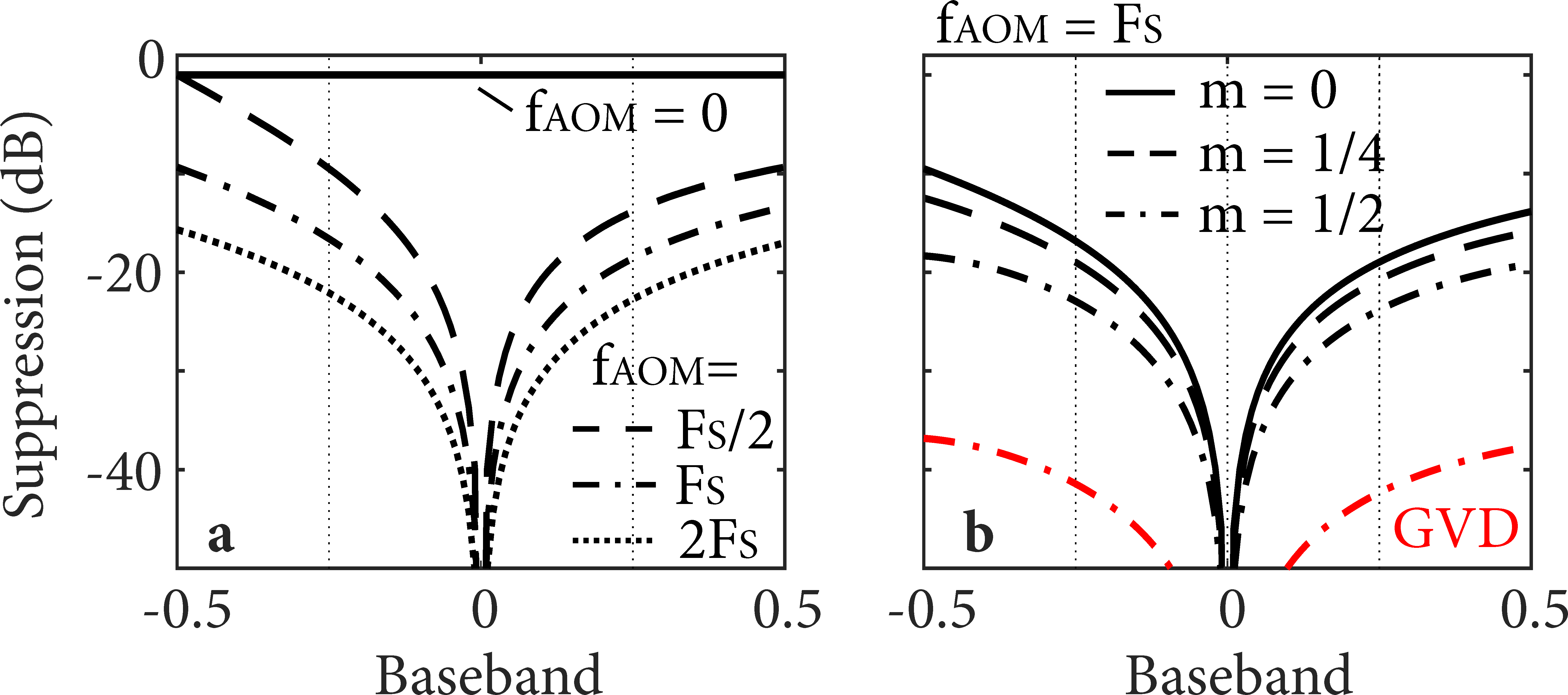}}
\caption{Simulated artifact suppression within 0th order baseband ($L_B$). (a) Artifact suppression for different frequency shifts $\pm f_{\rm{AOM}}$ corresponding to a depth translation of a fraction of the baseband depth. (b) Artifact suppression for a frequency shift of $F_s$ and different pulse modulation depths, $m$. The red curve takes into account the group velocity dispersion (GVD) of AMTIR ($\beta_2=820$~ps$^2$/km) and numerical dispersion compensation for $m=0.5$.}
\label{Fig7}
\end{figure}

The circular ranging method requires that the positive and negative delay spaces around the zero path delay are resolved independently, i.e., that the sign of the delay is measured in addition to the absolute value of the delay~\cite{Siddiqui}. In prior reports of circular ranging, this has been achieved by passive polarization-based optical quadrature demodulation~\cite{Siddiqui:0} and acousto-optic frequency shifting~\cite{Siddiqui}. 
Here, we adopt the acousto-optic frequency shifting (AOFS) approach. While straightforward to implement, the AOFS method can introduce complex conjugate artifacts that have not been described in detail. Here, we explain the origin of these signals, and discuss laser source and system design factors that can mitigate their impact. To this end, we first modeled and analyzed the interference fringes generated by a time-stepped frequency comb source. The model assumed a perfect frequency comb spectrum with a constant 80 GHz FSR.
In the time domain, each comb line is represented by a single finite pulse parameterized by the modulation depth $m$. The RF frequency content of the fringes was calculated for detection with either dual quadrature $I/Q$ outputs (i.e., like those generated by Refs.~\cite{Siddiqui:2,Lippok:2} and allowing discrimination of positive and negative RF frequencies), or with a single real output (i.e., the $I$ output without the $Q$ output which, as a result, does not distinguish positive and negative RF frequencies). All fringe simulations assume a single mirror reflection. In these simulations, the location of the mirror is parameterized by its circular delay rather than its physical delay; the circular delay is given by the physical delay modulo the circular delay range, $\Delta \tau = 1/\Delta \nu$ where $\Delta \nu$ is the FSR of the frequency comb. This corresponds to a circular depth range, $L_B=c\Delta\tau/2$. 

First, we examined the RF content with $I/Q$ outputs and without a AOFS [Fig.~\ref{Fig6}(a)]. Note that in circular ranging a single reflection generates a set of RF harmonics. These result from the discontinuous wavelength-stepping and amplitude pulsing of the source. A baseband signal ($0^{th}$ order) is located within the window defined from $-F_s/2$ to $+F_s/2$ where $F_s = 1/\Delta t=$~\emph{v}$/\Delta\nu_{\rm{FFP}}$. Higher order RF harmonics are located in adjacent, non-overlapping RF windows (i.e., +1st order from $F_s/2$ to $3F_s/2$). These higher RF orders contain the same information as the baseband signal and do not need to be captured for imaging. As such, detection and acquisition can be optimized for the baseband ($0^{th}$ order) window with a sampling rate (on each of the I and Q channels) of only $S=F_s = 1/\Delta t=2$\emph{v}$L_B/c$. With quadrature detection as shown in Fig.~\ref{Fig6}(a), low-pass analog filters can remove the higher order signals prior to digitization. 

Next, we investigated the use of an AOFS. Although the AOFS is used to avoid the need for I/Q detection, we first simulated the RF frequency response with both an AOFS and $I/Q$ detection [Fig.~\ref{Fig6}(b), upper panel] to more clearly illustrate the effect of the AOFS relative to the results of Fig.~\ref{Fig6}(a). As shown, the AOFS shifts the RF signals by the applied optical frequency shift, $f_{\rm{AOM}}$ in the RF domain. A frequency shift of $f_{\rm{AOM}} = F_s/2$ therefore shifts the baseband signals entirely into the positive RF frequency space. In principle, because these baseband signals of interest are located within the positive RF space, they can be captured by a single channel (I only). Figure~\ref{Fig6}(b) (lower panel) illustrates the RF content when detecting on a single channel. Here we can see that negative order harmonics (which are still located in the negative RF frequency space) generate artifacts in the baseband. Thus, while the AOFS is effective at placing the full range of baseband signals into an entirely positive (or negative) RF frequency space, this does not prevent the RF harmonics created by the frequency comb source from creating artifacts in the baseband. We note that with $f_{\rm{AOM}} = F_s/2$, the artifacts result from the $-1^{st}$ order harmonics. Higher values of $f_{\rm{AOM}}$ can be used to translate the baseband signals to higher RF frequencies and to increase the order of the harmonics that alias into the baseband window. Figure~\ref{Fig6}(c) illustrates imaging with $f_{\rm{AOM}} = F_s$, which selects the $-2^{nd}$ order harmonics. It is important to note that the RF harmonic amplitudes decay with order and the pulse shape influences the decay rate. Therefore, through manipulation of the pulse shape and the frequency shift, $f_{\rm{AOM}}$, the extinction between the baseband signal and these artifacts can be controlled. This is shown in Fig.~\ref{Fig7}(a) as a function of $f_{\rm{AOM}}$ for $m=0$ and in Fig.~\ref{Fig7}(b) as a function of $m$ for a fixed $f_{\rm{AOM}}=F_s$. The red curve in Fig.~\ref{Fig7}(b) shows the artifact suppression for $m=0.5$ while also taking into account the group velocity dispersion (GVD) of amorphous material (AMTIR) of a typical AOFS ($\beta_2=820$~ps$^2$/km at $\lambda_0=1.3~\upmu$m, 10 mm length). Numerical correction of dispersion for the baseband signals in the positive RF space causes broadening in the negative RF space, and further reduces the artifact peak. An artifact suppression of better than 30 dB can be achieved. We note that this broadening does not reduce the integrated power of the artifact signal. Higher values of $f_{\rm{AOM}}$ also increase the digitization rate and therefore force a compromise between reducing acquisition bandwidth (through lower values of $f_{\rm{AOM}}$) and reducing artifacts (through higher values of $f_{\rm{AOM}}$). 

\begin{figure}[!b]
\centerline{\includegraphics[width=7cm]{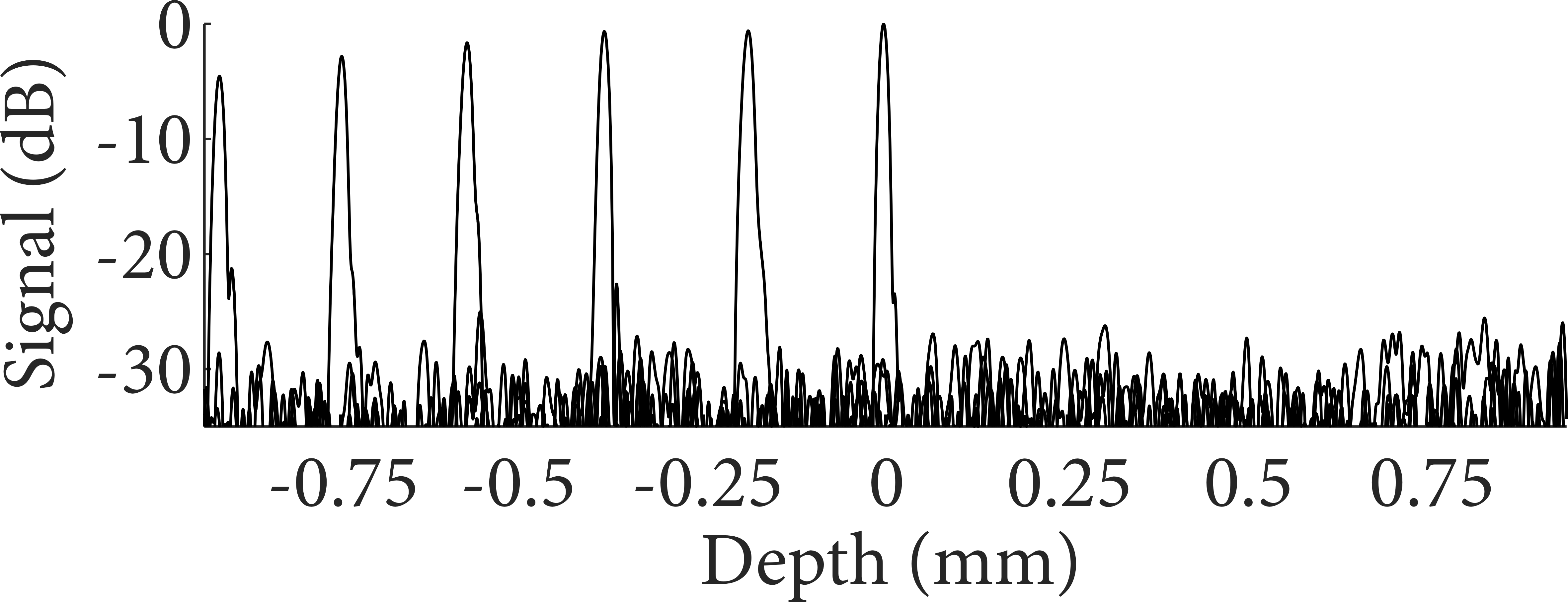}}
\caption{Plot showing the zero order baseband. Artifact suppression was better than 30~dB across the baseband.}
\label{Fig8}
\end{figure}

\section{Circular imaging}

The full imaging and data acquisition system is shown in Fig.~\ref{Fig1}. An AOFS with a frequency shift of 100 MHz (approximately $F_s$) was used. The laser provided output pulses described by $m\approx0.6$. We observed artifact suppression of more than 30 dB (Fig.~\ref{Fig8}). We note that this includes the influence of dispersion mismatch in the interferometer, which was compensated numerically. The PSF magnitude decays by approximately 4~dB towards the baseband edge. This decay rate depends on the pulse shape and pulse modulation depth as well as the circular depth range (FSR of FFP). This roll-off is stable and was compensated for numerically in generated images. Imaging was performed using lenses with focal lengths of 30~mm or 200~mm. The corresponding spot sizes were 25~$\upmu$m and 166~$\upmu$m, with confocal parameters of 751~$\upmu$m and 3.3~cm, respectively. A small portion of the light was used to generate a trigger signal using a fiber Bragg grating. A second interferometer provided a calibration signal to correct for sweep nonlinearities. Finally, signals were acquired at 400~MS/s using a data acquisition board. Although 319~MS/s was sufficient considering the AOFS with $S=2(vL_B/c+f_{\rm{AOM}})$, we chose a slightly higher sampling rate to also capture positive first order harmonic signals in the reconstructed tomograms. The imaging sensitivity ranged from 98~dB to 101~dB for single and dual SOA laser cavities, respectively.

\begin{figure}[!t]
\centerline{\includegraphics[width=\linewidth]{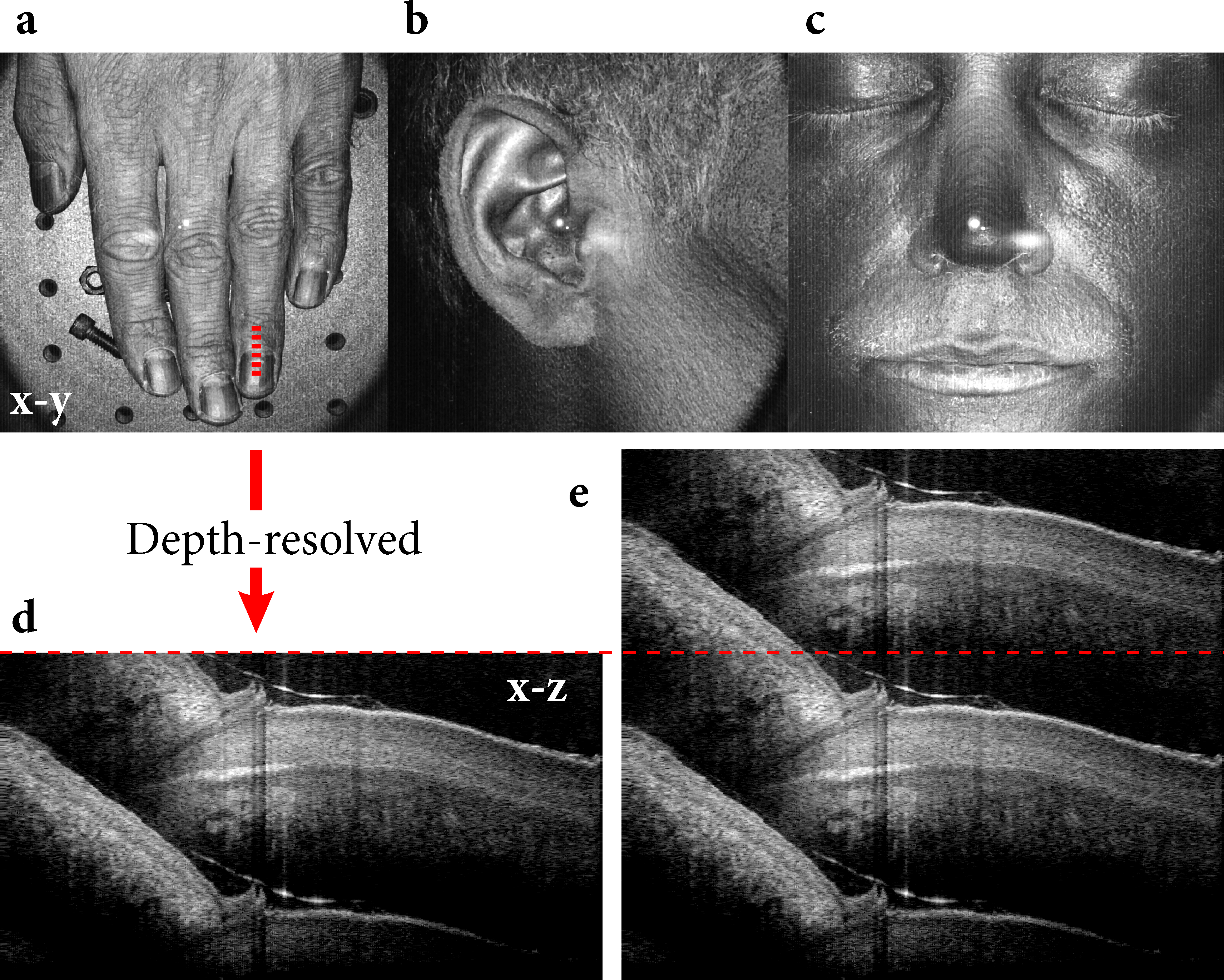}}
\caption{Volumetric projections (x-y) of a human hand (a), ear (b) and face (c) demonstrate the long-range imaging capability of the frequency comb source. The hand was positioned more than 5 cm above the optical table. (d) Depth-resolved, cross-sectional image (x-z) of a finger nail as illustrated by the red line in (a). (e) Superposition of two copies of the cross-sectional image seen in (d). The borderless wrapping of structure deeper than the baseband depth (red, dashed line) is obvious.}
\label{Fig9}
\end{figure}

A long-range imaging experiment is shown in Figure~\ref{Fig9}(a)-\ref{Fig9}(c) with wide field-of-view depth projections (x-y) of a human hand, ear and face. A hand was selected to generate a cross-sectional view to demonstrate that high imaging speed, range as well as axial resolution is maintained by the circular ranging technique. Figure~\ref{Fig9}(d) shows the depth-resolved view (x-z) along the finger nail as indicated by the red line in Fig.~\ref{Fig9}(a). The first order replica is clearly visible beneath the zero order baseband image. Superimposing a copy of the cross-section makes the borderless wrapping of scatterers outside the baseband clear [Fig.~\ref{Fig9}(e)]. 

\section{Discussion}

For traditional OFDI, imaging a range of 4~cm with a 488~kHz repetition rate and 105~nm sweep range requires a sampling rate of 4.7~GS/s and yields a data volume of 39~GB (assuming 1000~$\times$~1000 volume, 14~bit, $\lambda_0$ = 1310~nm). This is faster than current state-of-the-art acquisition boards. The use of circular ranging as shown in this work reduced this 15-fold (319~MS/s and 2.6~GB). The 1.88~mm circular depth range produced by the 80~GHz FFP FSR can capture the full detectable signal for most biological samples. A shorter circular depth range (i.e., larger FFP FSR) may be adequate for some samples, and would increase the data compression. Reducing self-phase modulation in the SOA is likely to improve the coherence length and thereby further extend imaging range. Additional intracavity dispersion compensation would enhance the FDML condition across the entire sweep range and may further extend the coherence length. Choosing a lower FFP finesse could improve laser noise performance as is indicative from the numerical simulation. The imaging speed was limited by the mechanical resonance frequency of the tunable spectral filter. Utilizing different filters or higher harmonics of the filter's mechanical resonance frequency will allow higher repetition rates. Current acousto-optic frequency shifters support repetition rates of several megahertz. For example, a repetition rate of 2~MHz requires a frequency shift of approximately $F_S=400$~MHz (assuming $\lambda_0$ = 1310~nm, $\Delta\lambda$ = 100~nm, $\Delta\nu_{\mathrm{FFP}}$ = 80~GHz, $v$ = 35~THz/$\upmu$s). Higher frequency shifts are available but have low diffraction efficiencies. 

\section*{ACKNOWLEDGMENTS}
This work was supported in part by the National Institute of Health (NIH) grant P41EB-015903.

\appendix
\section{Numerical simulation}

We describe here the model we used to simulate a stepped frequency comb laser in section III. For our analysis we consider a fiber ring cavity that includes a SOA, a tunable spectral bandpass filter and a coupler for optical feedback and output coupling. A frequency comb is generated considering a bulk Fabry-P\'erot etalon. The electric field inside a SOA is described by the coupled differential equations~\cite{Bilenca,Girard,Slepneva,Agrawal,Shtaif}
\begin{equation}
\frac{dE}{dz}+\frac{1}{v_g}\frac{dE(\tau,z)}{d\tau}=\frac{1}{2}[(1+i\alpha_H)g-a_{\rm SOA}]E+N_c(\tau,z)\,,\tag{A1}
\label{ES1}
\end{equation}

\begin{equation}
\frac{dh}{d\tau}=\frac{g_0l-h}{\tau_c}-\frac{|E(\tau)|^2}{E_{sat}}[\exp(h(\tau))-1]\,,\tag{A2}
\label{ES2}
\end{equation}
where the carrier induced index change is accounted through the linewidth enhancement factor $\alpha_H$, $g_0$ is the small signal gain, $\tau_c$ is the carrier life-time, $a_{\rm SOA}$ are the SOA losses, $E_{sat}=P_{sat}\tau_c$ is the saturation energy with $P_{sat}$ the saturation power, $v_g$ is the group velocity inside the SOA, $N_c$ represents the addition of a noise spectrum due to amplified spontaneous emission (ASE) and $\tau$ is the time reference frame. Furthermore, $h(\tau)$ is the time-dependent gain over SOA length $l$, $h(\tau)=\int^l_0g(z,\tau)dz$, and represents the integrated gain at each point of the temporal profile. Equation~\ref{ES1} was solved using the split-step Fourier method where the field propagation with group velocity $v_g$ was accounted for in the frequency-domain and the SOA gain (right hand side of Eq.~\ref{ES1}) was computed in the time-domain by solving Eq.~\ref{ES2} numerically. Moreover, the solutions for the SOA output were constrained by the boundary condition
\begin{equation}
\begin{split}
E(\omega,z=0)=E(\omega,z=l)\sqrt{\eta(1-\alpha_{\rm c})}F_{\mathrm{TFP}}(\omega)\\
\qquad \times F_{\mathrm{FFP}}(\omega)\exp[-i\beta(\omega)L]\,,
\end{split}\tag{A3}
\end{equation}
where $E(z=0)$ is the electric field at the SOA input, $E(z=l)$ ist the field at the SOA output, $\eta$ is the cavity feedback, $\alpha_{\rm c}$ are the cavity losses, $L$ is the cavity length, $\omega=2\pi\nu$ is the angular frequency and $\beta(\omega)$ is the fiber propagation constant that can be expressed as a Taylor series expansion around a center frequency, $\beta(\omega)=\sum_{{\rm k}\geq2}^\infty\frac{\beta_{\rm k}}{{\rm k}!}(\omega-\omega_0)^{\rm k}$, with $\beta_{\rm k}=(d^{\rm k}\beta/d\omega^{\rm k})_{\omega=\omega_0}$, where we only consider second order chromatic dispersion. Furthermore, $F_{\mathrm{TFP}}(\omega)=\left|1/[1+i2\omega/(2\pi\delta\nu_{\mathrm{TFP}})]^2\right|$ is the tunable optical bandpass filter with a filter bandwidth $\delta\nu_{\mathrm{TFP}}$ (FWHM). The transmission through the fixed etalon was accounted for by $F_{\mathrm{FFP}}=1/[1+F_{\mathrm{FFP}}\sin^2(\delta/2)]$, with $\delta=\omega2n_{\mathrm{FFP}}L_{\mathrm{FFP}}\cos(\theta)/c$. The propagation angle, $\theta$, was assumed zero and the etalon length is given by ${L_{\mathrm{FFP}}=c/(2n_{\mathrm{FFP}}\Delta\nu_{\mathrm{FFP}})}$, with $\Delta\nu_{\mathrm{FFP}}$ being the FSR and $n_{\mathrm{FFP}}$ the etalon refractive index. The simulations were performed in the swept filter reference frame starting at a center wavelength of 1310 nm. Starting from spontaneous emission noise, we computed iteratively the gain and the optical field until a steady-state regime was observed. Unless stated in the text or figures, the following values were chosen for the simulations: $L=3$~m, $G_0=\exp(g_0l)=29$~dB, $E_{sat}= 7.92$~pJ, $v_g=8.4272\times10^7$~m/s, $\alpha_H=5$, $\tau_c=440$~ps, $a_{c}=0.6$, $\eta=0.1$, $\beta_2=1$~ps$^2$/km, $\delta\nu_{\mathrm{TFP}}=40$~GHz, $\Delta\nu_{\mathrm{FFP}}=80~$GHz, $F_{\mathrm{FFP}}=80$, $n_{\mathrm{FFP}}=1.45$.

\end{document}